\documentclass[11pt]{article}

\usepackage[margin=1in]{geometry}
\usepackage{graphicx}
\usepackage{amsmath}
\usepackage{booktabs}
\usepackage{hyperref}
\usepackage[numbers,sort&compress]{natbib}
\usepackage{setspace}

\graphicspath{{./}{../../manuscript_arxiv/src/}}

\title{Low-Cost Continuous-Wave Diffusive Microtomography with Fiber-Scanned White-Light Illumination}
\author{Alexander Ingold\\Department of Electrical and Computer Engineering, University of Utah}
\date{}

\hypersetup{
  pdftitle={Low-Cost Continuous-Wave Diffusive Microtomography with Fiber-Scanned White-Light Illumination},
  pdfauthor={Alexander Ingold}
}

\newcommand{\autocite}[1]{\cite{#1}}

\begin{document}
\maketitle

\begin{abstract}
Tomographic microscopy enables three-dimensional internal imaging but often requires expensive optical or X-ray instrumentation. Here we present an ultra-low-cost continuous-wave diffusive tomography (CWDT) system for biological samples. The system uses a smartphone microscope, a white LED coupled into an optical fiber, 3D-printed micropositioners, and a physics-based forward model optimized with machine learning. We demonstrate full-color volumetric reconstructions from a tartrazine-cleared poplar section, a scattering phantom, fungal mycelium near an \textit{Arabidopsis} root, and thick poplar branch imaging with an inserted side-emitting fiber. The current results are qualitative and exploratory, but they show that scanned fiber illumination and inexpensive hardware can produce useful three-dimensional reconstruction outputs for low-cost microscopy experiments.
\end{abstract}

\section{Introduction}

Tomographic microscopy, which enables 3D internal imaging of specimens, is a powerful but often prohibitively expensive capability. Key modalities such as optical coherence tomography and two-photon microscopy rely on costly components including ultrafast pulsed lasers and high-speed detectors. Similarly, micro X-ray computed tomography requires specialized X-ray sources and sensitive detector arrays. This capital investment limits use cases such as long-term field deployment, point-of-care medical diagnostics in low-resource settings, and hobbyist microscopy.

Our goal is to create an ultra-low-cost micro-tomography system for imaging 3D volumes in biological samples. We utilize a novel machine-learning-based variation of continuous wave diffusive tomography (CWDT) \autocite{Guan2025}. Our primary innovations are the use of a physics-based forward model optimized with machine learning, combined with white-light illumination and a scattering-reduction dye to enable high-resolution, full-color volumetric imaging with inexpensive hardware. CWDT involves positioning a continuous wave, fiber-guided, LED light source in an array of positions behind the sample and imaging at each position. CWDT has been explored for breast cancer diagnosis \autocite{Culver2003} and functional neuro-imaging \autocite{Galderisi2016}.

While many CWDT projects use near-infrared illumination to avoid the absorbance of hemoglobin in animal tissues, we chose white-light illumination because it aligns with the low-cost goal of the project and enables imaging and reconstruction of three-color volumes. To improve image quality, the sample is soaked in a strongly absorbing tartrazine dye, which can decrease light scattering within tissue \autocite{Zihao2024}. This non-toxic, food-safe dye shifts the real and complex refractive-index components. The real part helps match tissue interfaces to reduce scattering, while the imaginary part increases absorption to suppress multiple scattering events, improving the signal-to-noise ratio and reducing the ill-posedness of the inverse problem.

The resulting stack of illuminations is processed into a 3D physics and ray-tracing-based forward model of the tissue. Light is simulated to radiate from a single illumination point and travel through approximately 16 absorbance and scattering filters.

We imaged a 940 $\mu$m x 940 $\mu$m x 250 $\mu$m volume of a tissue section with a mean squared error of 0.002. We modeled 20 depth slices for a z-axis sampling interval of 10 $\mu$m, an x/y-axis resolution of 1.5 $\mu$m on the surface, and approximately 50--100 $\mu$m resolution at the deepest depth.

To demonstrate the versatility of this technique beyond prepared tissue sections, we applied it to the challenging task of in-vivo imaging of thick, living plant matter. For thick poplar branches, the illumination source is inserted into the sample rather than positioned beneath a thin section. The optical fiber is ground at a 45-degree angle, and the end is treated with reflective paint or metal sputter coating. The mirrored fiber directs the light out of the end of the fiber at a 90-degree angle \autocite{davenport2016method}. This protocol enables imaging of living poplar branches still attached to the tree. Imaging thick samples while maintaining high resolution using optical fiber insertion could also be relevant to tomographic imaging of precancerous skin lesions.

\begin{figure}
    \centering
    \includegraphics[width=1\linewidth]{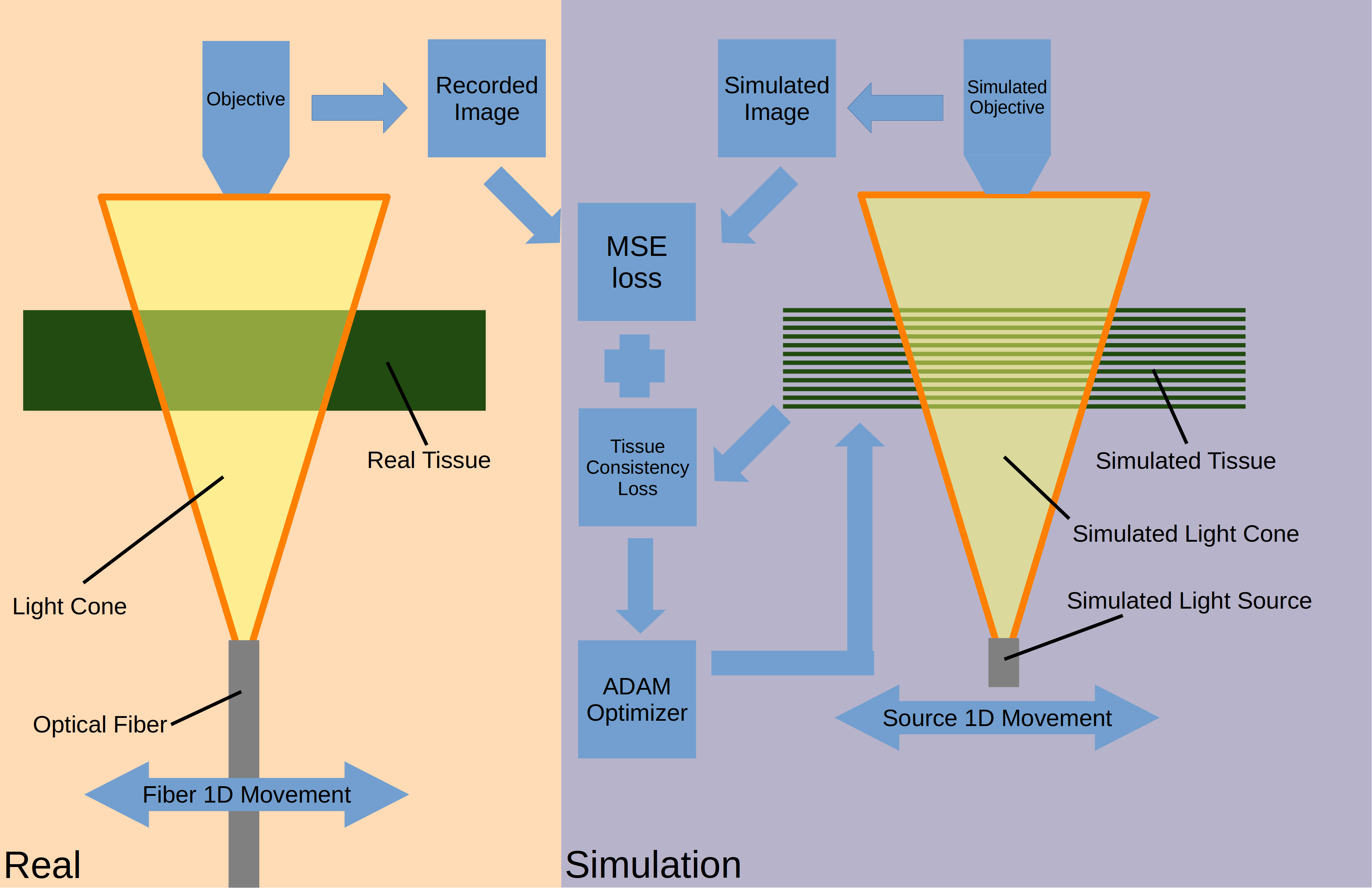}
    \caption{A simulated physics-based tissue model with learned absorbance parameters and a simulated light source approximates the sample images with multiple incident light source locations.}
    \label{fig:schematic}
\end{figure}

The overall CWDT image-formation model is summarized in Figure~\ref{fig:schematic}.

\section{Methods}
\subsection{Microscope}
Diffusive tomography requires a movable transmission light source and a camera imaging the surface of the sample. The light source consisted of a white LED flashlight (BLITZU Gator 32) and an optical fiber (Thorlabs FG050LGA, 50 $\mu$m core, 125 $\mu$m cladding). One end of the optical fiber was positioned in direct contact with the LED. The other end of the optical fiber was positioned underneath thin sections or inserted into thick samples. Both ends of the fiber were positioned with TriAxis 3D-printed micropositioners \autocite{IngoldTriAxisMakerWorld,IngoldTriAxisThingiverse,IngoldTriAxisOnshape}.

The microscope was a low-cost 120x smartphone-based microscope. A 3D-printed frame supported the microscope, smartphone (Motorola ThinkPhone 50 MP camera), microscope slide, and two 3D-printed micropositioners for the light source in inserted or transmission positions. The light source was moved in 100 $\mu$m increments between image captures. Camera settings were adjusted for each sample, but focus, ISO, exposure time, and white balance were held constant across each image stack.

\subsection{Optical Fiber Preparation}
For transmission imaging, the optical fiber was cut with scissors. For thick-sample imaging, the end of the optical fiber was modified so that light exits perpendicular to the fiber axis. The fiber was ground at a 45-degree angle with lapping sheets (Thorlabs LF5P Silicon Carbide 5 $\mu$m Grit). For rapid grinding, a lapping-sheet disk was mounted on a rotary grinder (Dremel 1.2A), and the fiber was held against the lapping sheet with a 3D-printed micropositioner for 10 minutes. Lapping can also be performed by hand. After lapping and inspection for defects with the miniature microscope, the end of the fiber was dipped in mirror-finish paint (Rust-Oleum Mirror Effect Spray Paint). This paint consists of reflective aluminum microparticles and a clear binder. Three to five dip coats were applied for complete coverage.

\subsection{Poplar Sample Preparation}
Branch cuttings from one-year-old \textit{Populus nigra} x \textit{deltoides} trees were potted in Sunshine Mix \#4 (Sun Gro Horticulture) and maintained in a greenhouse for two years under a 14 h light / 10 h dark photoperiod. Imaging samples were prepared from branches at varying developmental stages, specifically between the 5th and 10th internodes. A narrow strip of bark ($\sim$3 cm in length, $\sim$200 $\mu$m in depth) was carefully removed using a miniature wood plane (Veritas Miniature Bench Plane, Lee Valley Tools) to expose the underlying xylem. For thick-sample imaging, the branch was clipped and used directly. For thin-section imaging, a second pass with the sample plane removed a $\sim$200 $\mu$m thick tissue section. 

A 0.38 molar tartrazine solution (FD\&C Yellow 5, sciencekitstore.com) was prepared by dissolving 10 grams of tartrazine in 50 ml of distilled water. Stirring and heating to 50 degrees Celsius were required to complete the dissolution. For tartrazine staining, samples were soaked overnight for 18 hours, patted dry, and mounted on a slide. Branch samples were fixed to the slide with quick-cure two-part epoxy, and the incision was covered with a thin layer of glycerol to preserve tissue clarity and prevent air infiltration. Thin sections were mounted on the slide with a coverslip and glycerol.

\subsection{Phantom Preparation}
A phantom was prepared for testing the resolution of suspended particles at various depths and sizes. The substrate was clear silicone sealant (JB Weld Clear RTV silicone). The suspended particles were silicon carbide abrasive (MJ Tumblers). The particles were sieved to a 1:1 mix of 120/220 grit and 400/600 grit, corresponding approximately to particle sizes of 65--125 $\mu$m and 25--45 $\mu$m. Cow's milk was added to increase scattering at a concentration of 10\% by volume.

\subsection{Imaging}

For thick samples such as the branch of a living poplar tree, the optical fiber was positioned approximately 1 mm below the surface of the branch. After removing a 5 mm region of bark, a hypodermic needle was inserted parallel to the plane of the incision and perpendicular to the branch. The optical fiber was threaded through the needle, and the needle was then removed, leaving only the optical fiber. The fiber was rotationally aligned to produce a vertical beam, lubricated with glycerol, and pulled through the branch in 100 $\mu$m steps with a 3D-printed micropositioner.

For thin samples, the optical fiber is positioned underneath and normal to the sample. The distance from the fiber to the slide is set to 4 mm using the 3D-printed micro-positioner's z-axis, aligned to the center of the field of view with the y-axis, and translated with 100 $\mu$m steps with the x-axis. 

\subsection{Computational Methods}

Continuous-wave diffusive tomography transforms a stack of images captured with different transmission-illumination positions into a multilayer 3D model of the sample. The physics-based model simulates a scattering medium with a spatially varying attenuation parameter and three color channels.

\noindent\textbf{Input processing:} The image stack is cropped to the microscope's field of view and rotated to align the light-source movement with the x-axis. Optional bilinear interpolation can downscale the image stack to reduce computation time.

\noindent\textbf{Light-source modeling:} The Gaussian-patterned illumination source is modeled at the same measured positions as the real fiber illumination source. The light source can be adjusted to inverse-square, Gaussian, multimodal-fiber, or combined Gaussian and multimodal modes. The multimodal-fiber mode simulates the illumination as a cone, with numerical aperture determining the cone angle, while the Gaussian method is recommended for most applications. An interactive GUI helps the user adjust light-source parameters such as x, y, and z position; x step; step direction; intensity; light distribution; light size; and fiber numerical aperture. Accurate microscope micron-to-pixel conversion and sample thickness are important for reliable modeling. Capturing an image stack without the sample is recommended, but not required, for adjusting light-source parameters. The light-source position is modeled as

\begin{equation}
\begin{pmatrix}
x_{\text{src}} \\
z_{\text{src}}
\end{pmatrix} = \begin{pmatrix}
\frac{N_{\text{pix}}}{2} + \frac{(i - i_{\text{center}}) \cdot s_{\text{step}} \cdot d \cdot N_{\text{pix}}}{\text{FOV}_{\mu m}} \\
-\frac{z_{\text{depth}} \cdot 1000 \cdot N_{\text{pix}}}{\text{FOV}_{\mu m}}
\end{pmatrix}
\label{eq:source_coords}
\end{equation}

\noindent where $i$ is the image index, $i_{\text{center}}$ is the center image, $s_{\text{step}}$ is the step size in mm, $d$ is the direction ($\pm 1$), $N_{\text{pix}}$ is the image size in pixels, and $\text{FOV}_{\mu m}$ is the field of view in microns.

\noindent\textbf{Ray-tracing model:} A geometric light-transport model projects the light source through the attenuation layers. During ray tracing, the simulated light source is projected through a series of attenuation layers, with the image projected along rays originating from the light source at each layer. Equation~\ref{eq:ray_shift} shows how light is transported to the next layer, where bilinear interpolation is used for sub-pixel sampling during ray projection in a normalized coordinate system. Padding is added to lower layers to facilitate off-axis illumination. The sample is approximated with uniform scattering applied at each layer with a Gaussian kernel, because a spatially varying scatter parameter did not significantly increase performance.

\begin{equation}
\begin{pmatrix}
x_{\text{out}} \\
y_{\text{out}}
\end{pmatrix} = \frac{D_{\text{layer}}}{z_{\text{src}}} \begin{pmatrix}
\text{row}_{\text{in}} - x_{\text{src}} \\
\text{col}_{\text{in}} - y_{\text{src}}
\end{pmatrix}
\label{eq:ray_shift}
\end{equation}

\noindent where $D_{\text{layer}}$ is the distance between consecutive attenuation layers in pixels, $z_{\text{src}}$ is the z-coordinate of the light source as a negative depth in pixels, $(x_{\text{src}}, y_{\text{src}})$ are the light-source coordinates in the image plane, $(\text{row}_{\text{in}}, \text{col}_{\text{in}})$ are the input pixel coordinates, and $(x_{\text{out}}, y_{\text{out}})$ are the corresponding pixel-shift offsets applied during ray projection to the next layer.

\begin{equation}
\mathbf{I}_{\text{out}}^{(c)} = \mathcal{R}\left( \mathcal{G}_\sigma \left( \mathbf{I}_{\text{in}}^{(c)} \odot \alpha_L^{(c)} \right), \mathbf{s}, D_L \right)
\label{eq:forward_model}
\end{equation}
\noindent where $\mathcal{R}$ is the ray-tracing operator in Equation~\ref{eq:ray_shift}, $\mathcal{G}_\sigma$ is Gaussian scattering with parameter $\sigma$, $\odot$ is element-wise multiplication, $c$ indexes color channels, and $L$ indexes layers.

\noindent\textbf{Optimization:} The spatial pattern of the attenuation layers is optimized with the Adam optimizer and a mean squared error loss function until the output of the physics-based simulation matches the microscope images. The PyTorch framework handles gradient computation and CUDA tensor-based GPU acceleration. Optional learning-rate decay can be activated through configuration parameters. Additional loss functions encourage consistency across layers in the average attenuation value and in the variance of the Laplacian sharpness value. To accelerate optimization, the absorbance values are initialized from the recorded image with central illumination. The initial values are scaled by taking the nth root, where n is the number of absorbance layers.

\begin{equation}
\mathcal{L}_{\text{total}} = \text{MSE}(\mathbf{I}_{\text{pred}}, \mathbf{I}_{\text{target}}) + 5.0 \cdot \mathcal{L}_{\text{pixel}} + 20.0 \cdot \mathcal{L}_{\text{laplacian}}
\label{eq:total_loss}
\end{equation}

\begin{align}
\mathcal{L}_{\text{pixel}} &= \frac{1}{C} \sum_{c=1}^{C} \frac{1}{L} \sum_{l=1}^{L} \left| \bar{\alpha}_{l,c} - \frac{1}{L} \sum_{l'=1}^{L} \bar{\alpha}_{l',c} \right| \label{eq:pixel_consistency} \\
\mathcal{L}_{\text{laplacian}} &= \frac{1}{C} \sum_{c=1}^{C} \frac{1}{L} \sum_{l=1}^{L} \left| \text{Var}(\nabla^2 \alpha_{l,c}) - \frac{1}{L} \sum_{l'=1}^{L} \text{Var}(\nabla^2 \alpha_{l',c}) \right| \label{eq:laplacian_consistency}
\end{align}

\noindent where $\mathcal{L}_{\text{total}}$ is the combined loss function, $\mathbf{I}_{\text{pred}}$ and $\mathbf{I}_{\text{target}}$ are the predicted and target RGB images, $\mathcal{L}_{\text{pixel}}$ enforces similar average attenuation values across layers, $\mathcal{L}_{\text{laplacian}}$ promotes consistent spatial detail across layers, $C$ is the number of color channels, $L$ is the number of layers, $c$ and $l$ are channel and layer indices, $\bar{\alpha}_{l,c}$ is the spatial average attenuation for layer $l$ and channel $c$, and $\text{Var}(\nabla^2 \alpha_{l,c})$ is the variance of the Laplacian measuring spatial sharpness in each layer.

\noindent\textbf{Confidence analysis:} Confidence analysis of the 3D model is critical when no ground-truth volume is available. It is impractical and inconsistent with the ultra-low-cost goal to use expensive systems such as two-photon microscopy or micro-computed X-ray tomography for every sample. Instead, we assess confidence with two methods. For confidence in the x and y axes, an x/y map of absolute error is generated by subtracting the simulated light output from the real recorded image for each illumination position and then averaging the error maps across all positions. For confidence in the z axis, we use Fourier-transform resolution analysis to assess the highest recoverable spatial frequency. For each image, we computed the 2D fast Fourier transform and power spectrum. The spectrum was azimuthally averaged into a 1D radial profile for robust quantification. The resolution cutoff was defined as the frequency where the log radial power dropped below 0.1, marking the transition from signal-dominated to noise-dominated content. This cutoff was converted to real-space resolution using pixel size and field of view, yielding a reproducible measure of image sharpness.

\noindent\textbf{Implementation and output:} A JSON configuration file contains all necessary parameters for hyperparameter optimization. The script automatically saves comprehensive outputs, including attenuation layers, error maps, target images, simulated incident light, simulated light at every layer and illumination position, predicted images for illumination layers, x/y error maps, z-axis Fourier resolution analysis, loss logs, cropping parameters, source code, and configuration files.

After computation, the tomographic model can be visualized in 3D space. The 3D visualization shows the absorbance values as a heat map that can be rotated and flown through. 

In addition to the above computational methods, we explored a Gaussian-splatting model and a convolutional U-Net model. These methods did not perform as well as the main light-transport model. The methods and results are described in the supplement.

\section{Results}
\subsection{Reconstruction Examples}

\begin{figure}
    \centering
    \includegraphics[width=1\linewidth]{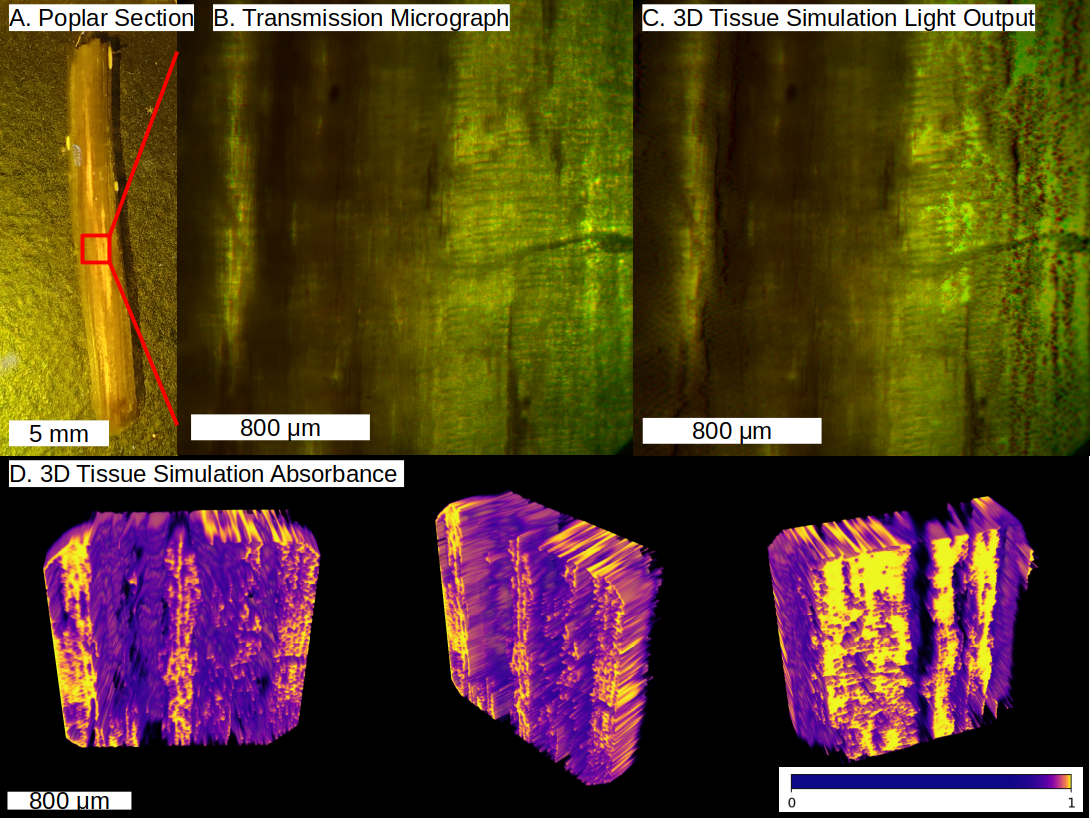}
    \caption{A. A 250 $\mu$m thick tangential section of poplar branch was soaked for 18 hours in tartrazine. B. Transmission images were captured. C. The simulated light output from the tomographic reconstruction of the poplar section. D. Absorbance values rendered in a heat map from three views: front, angled, and rear. }
    \label{fig:transmission}
\end{figure}

\begin{figure}
    \centering
    \includegraphics[width=1\linewidth]{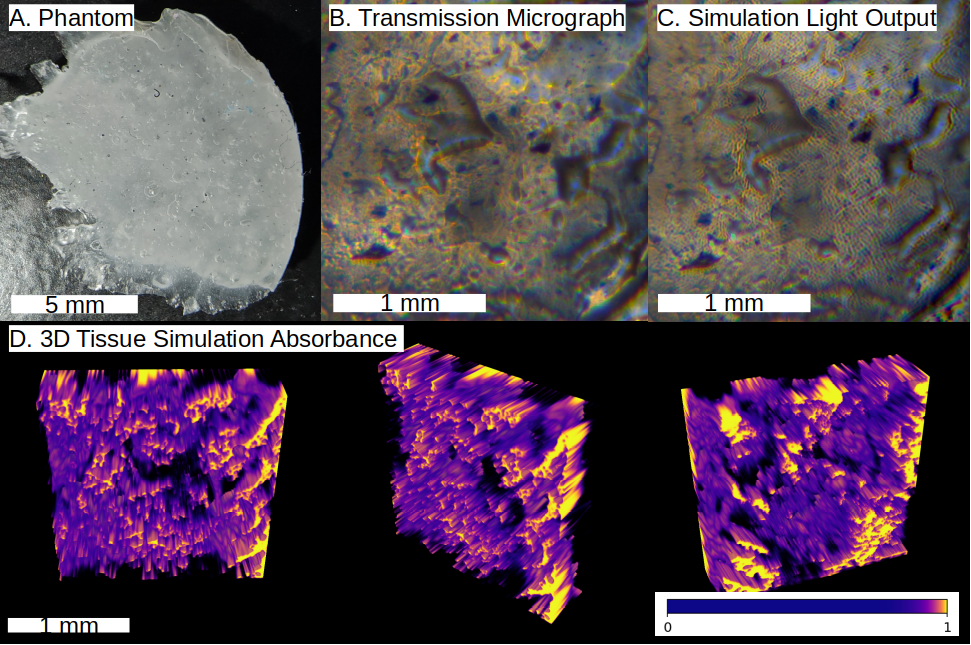}
    \caption{A. A 2 mm thick section of an optical phantom made of silicon carbide sieved for 25--125 $\mu$m particles, milk, and clear silicone sealant. B. Transmission images were captured. C. The simulated light output from the tomographic reconstruction of the phantom section. D. Absorbance values rendered in a heat map from three views: front, angled, and rear.}
    \label{fig:phantom}
\end{figure}

\begin{figure}
    \centering
    \includegraphics[width=1\linewidth]{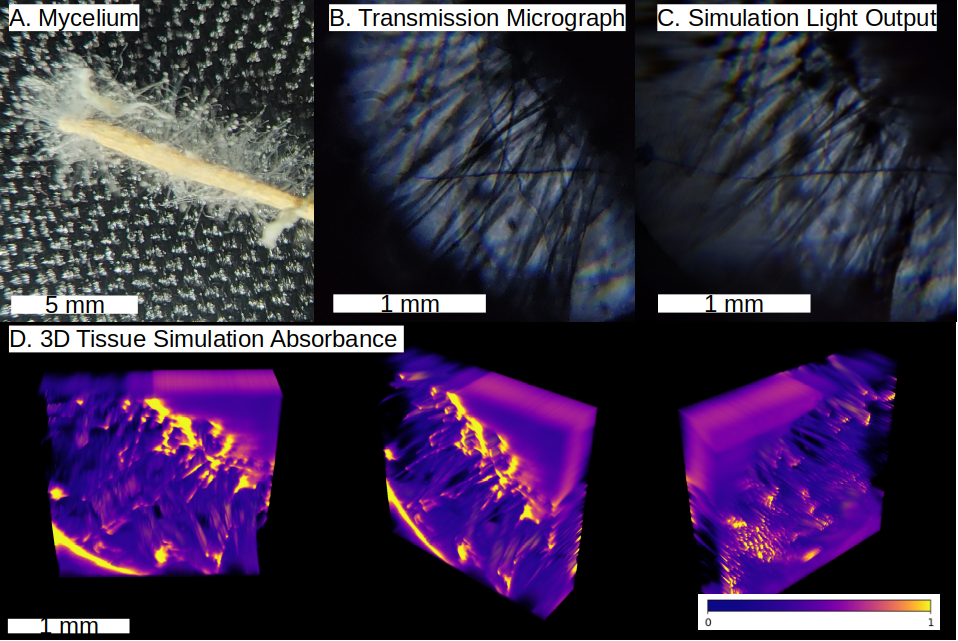}
    \caption{A. Fungal mycelium growing from Arabidopsis root. B. Transmission images were captured with varying light positions. C. The simulated light output from the tomographic reconstruction of the fungal sample. D. Absorbance values rendered in a heat map from three views: front, angled, and rear.}
    \label{fig:fungus}
\end{figure}

\begin{figure}
    \centering
    \includegraphics[width=1\linewidth]{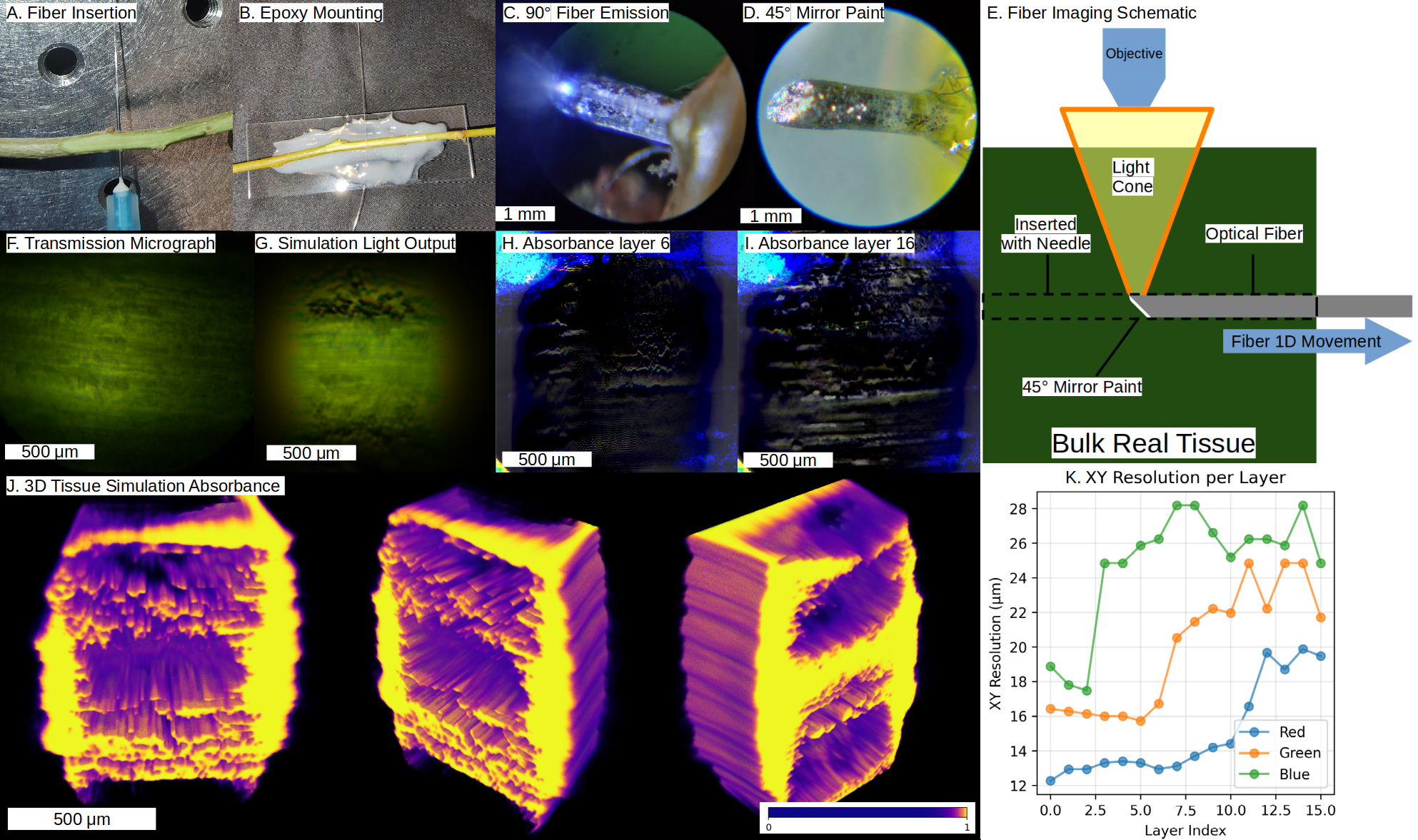}
    \caption{A. A 90 degree emission optical fiber is inserted into 3.5 mm thick poplar branch with a hypodermic needle. B. The poplar branch is fixed to a slide with epoxy. C. Alignment image of the 90 degree emission optical fiber before pulling through the sample. D. The 90 degree emission optical fiber is rotated to view the 45 degree ground bevel and mirror paint. E. Schematic of the fiber insertion tomographic imaging protocol. F. Transmission images captured. G. The simulated light output from the tomographic reconstruction of the poplar sample. H. 3-color Absorbance map at layer 6, black is high absorbance. I. 3-color absorbance map at layer 16, the most superficial layer. Black is high absorbance. J. Absorbance values rendered in a heat map from three views: front, angled, and rear. K. Resolution vs simulated tissue layer, analyzed with Fourier transform. }
    \label{fig:insertion}
\end{figure}

The reconstruction examples cover a tartrazine-cleared poplar section (Figure~\ref{fig:transmission}), a scattering phantom (Figure~\ref{fig:phantom}), fungal mycelium near an Arabidopsis root (Figure~\ref{fig:fungus}), and thick-branch imaging with an inserted side-emitting fiber (Figure~\ref{fig:insertion}).

\section{Discussion}

The results demonstrate a low-cost CWDT workflow that combines movable fiber illumination, smartphone-based microscopy, and a learned physics-based reconstruction model. The strongest case is the thin tartrazine-treated poplar section, where transmission geometry is simple and scattering is reduced. The phantom, fungal sample, and inserted-fiber branch experiments show that the same framework can be applied to more challenging samples.

The main limitation is that these reconstructions are not yet independently validated three-dimensional ground truth measurements. The model is optimized to reproduce the measured image stack, so the reconstructed volumes should be interpreted as qualitative absorbance models. Future work should compare the reconstructions with independent volumetric imaging, improve the light-transport model for thick scattering samples, and test whether the approach can support repeated imaging of living tissue.


\section*{Code Availability}
Reproducible code and configuration files for the Figure~2--5 reconstructions are prepared for release at \url{https://github.com/theMenonlab/diffusive_tomography}.

\section*{Hardware Availability}
Printable TriAxis micropositioner files are included with the code release at \url{https://github.com/theMenonlab/diffusive_tomography}, including STL files and the 3MF print project. Public mirrors are available through MakerWorld, Thingiverse, and the editable Onshape CAD document \cite{IngoldTriAxisMakerWorld,IngoldTriAxisThingiverse,IngoldTriAxisOnshape}.

\section*{Data Availability}
Figure-level raw inputs, model configurations, selected outputs, and manifests are available through the Kaggle dataset \url{https://www.kaggle.com/datasets/alingold/continuous-wave-diffusive-tomography}.

\section*{Acknowledgments}
The author thanks Rajesh Menon and the University of Utah Department of Electrical and Computer Engineering for guidance, laboratory support, and infrastructure. Funding from U.S. Department of Energy grant 55801063 is gratefully acknowledged.

\section*{Competing Interests}
The author declares no competing interests.

\appendix
\renewcommand{\thefigure}{S\arabic{figure}}
\renewcommand{\thetable}{S\arabic{table}}
\renewcommand{\theHfigure}{S\arabic{figure}}
\renewcommand{\theHtable}{S\arabic{table}}
\setcounter{figure}{0}
\setcounter{table}{0}
\section{Supplementary Material}
\section{Tartrazine for Deep Tissue Imaging in Poplar}
This supplement describes exploratory tartrazine experiments, an alternate Gaussian-splatting model, the relevant parts list, and the sample-preparation and imaging procedures used for the CWDT tartrazine study.

\subsection{Introduction}

A key objective in our research is deep tissue imaging of living cells within poplar trees. Living cells that are actively growing and dividing are concentrated in meristematic tissues of the cambial region near the xylem-phloem boundary \autocite{LINDSAY202450,Du2023PoplarMeristem}. While optical-fiber-based probe microscopes are an effective and minimally invasive method, an incision is required to reach deeper tissue layers \autocite{Ingold:24}. While mechanical incisions can reach the cambial region, they expose living cells, leading to drying, cell death, and wound response. This supplemental section explores a chemical method of tissue clarification to see deeper within tissue without a mechanical incision. More realistically, we might visualize living cells slightly deeper than the incision and therefore better protect the living cells.

Recently, a paper in \textit{Science} explored the use of the food-safe dye tartrazine for tissue clarification in animals \autocite{doi:10.1126/science.adm6869}. Tartrazine strongly absorbs wavelengths between 300 and 500 nm while reflecting red, yellow, and orange. Tartrazine increases the real part of the refractive index for red wavelengths without increasing absorption, according to the Kramers--Kronig relations. We aimed to replicate the results with a chicken-breast sample and determine whether tartrazine's tissue-clarification properties extend to plant tissue. Next, we determined whether tartrazine increases the depth of our existing ultraviolet (UV) autofluorescence microscopy techniques. Finally, we explored the effects of tartrazine on transmission microscopy.

\subsection{Methods}
The methods below describe sample preparation, graph-paper imaging, miniscope acquisition, tartrazine reflectance testing, and fiber-transmission measurements.

\subsubsection{Sample Preparation}

A 0.38 molar tartrazine solution (FD\&C Yellow 5, sciencekitstore.com) was prepared by dissolving 10 grams of tartrazine in 50 ml of distilled water. Stirring and heating to 50 degrees Celsius were required to complete the dissolution.

For Figure~\ref{fig:cwdt-supp-plant}, thin slices of poplar branch tangential to the surface, approximately 200 microns in thickness, were prepared for staining with a wood plane (Veritas Miniature Bench Plane). For Figure~\ref{fig:cwdt-supp-chicken}, chicken-breast samples were prepared by cutting a 1--4 mm thick section with a razor blade from the ventral side of the chicken breast, tangential to the surface. For Figure~\ref{fig:cwdt-supp-miniscope1}, a full-thickness sample of poplar wood was prepared by removing the bark of a poplar branch with a wood plane.

\subsubsection{Graph-Paper Imaging}

Before staining, the poplar and chicken sections were imaged on top of high-contrast graph paper to estimate the baseline bidirectional light transmission. The samples were submerged in 0.38 molar tartrazine for up to 18 hours. Upon removal from the tartrazine solution, the samples were padded dry with a paper towel and imaged again on top of the graph paper. The imaging was performed with a smartphone camera for Figures~\ref{fig:cwdt-supp-plant} and \ref{fig:cwdt-supp-chicken}.

\subsubsection{Miniscope}

Microscopic imaging for Figures~\ref{fig:cwdt-supp-miniscope1}, \ref{fig:cwdt-supp-transmission}, and \ref{fig:cwdt-supp-fiber} was performed using the UCLA miniscope. We used the UCLA Miniscope V4 (open-ephys.org) \autocite{open-ephys-miniscope,aharoni-lab-miniscope-v4}. The miniscope is modified for UV autofluorescence imaging of lignin, a major component of structural wood. The miniscope contains a Corning Varioptics electrowetting lens (A-25H0) 2.5 mm in diameter. This results in $\pm 200$ um of working-distance adjustment.

This system is modified for imaging lignin autofluorescence using UV excitation with a 390 nm LED (Lumileds, LHUV-0385-A045) and a 385/70 nm excitation filter (Chroma, ET385/70x), paired with a 455/50 nm emission filter (Chroma, ET455/50m) and a 412 nm long-pass dichroic mirror (Chroma, T412Ipxt). This setup overlaps the blue lignin autofluorescence band reported for UV-excited wood, including emission near 455--465 nm under 355 nm excitation \autocite{Donaldson2020}. The miniscope is controlled with the Miniscope DAQ V3.3 and Aharoni Lab Miniscope DAQ QT Software \autocite{AharoniLab2024MiniscopeDAQ}.

\subsubsection{Tartrazine Reflectance}

We investigated fluorescence imaging of tartrazine-soaked poplar branches because reflectance is more compatible with chronic observation of living cells than transmission through thin sections. Tartrazine reflectance imaging aims to answer the question: does tartrazine treatment increase the depth of imaging penetration into the tissue?

We noticed that the fluorescence level changes while the sample dries in air, so we present a tartrazine-soaked sample imaged after drying in air in Figure~\ref{fig:cwdt-supp-miniscope1}. Two sections of poplar branches were cut from the tree. The bark was removed with a wood plane, and three fiducials were made with a 0.25 mm punch. The samples were imaged with a dissection microscope and a miniscope. Next, the samples were soaked in tartrazine for 18 hours. The samples were removed from the tartrazine, padded dry with a lint-free cloth, and imaged with the miniscope. The fiducials were positioned in the bottom-left corner of the field of view. A second sample was covered with immersion oil and a coverslip to test whether covering the incision altered the drying response. Both samples were imaged again at 2 hours and 24 hours. Focus stacks were captured using the miniscope's electrowetting lens to keep the rough and curved branch surface in focus. The stacks were projected into a single image using a complex wavelet transform projection \autocite{CWT}.

\subsubsection{Fiber Transmission}

To take advantage of tartrazine's higher light transmission, we hypothesized that applying tartrazine in transmission microscopy could improve image quality over the non-tartrazine case. First, we tried transmission microscopy with sections of poplar soaked in tartrazine, and then we used an optical fiber for transmission microscopy of full-thickness branches, potentially preserving living cells.

Sections of poplar, tangential to the branch surface, were cut with a wood plane and measured with calipers to be 150--200 microns. The tartrazine section was soaked overnight in 0.375 molar tartrazine solution, while the non-tartrazine section was cut fresh from the branch. Both sections were mounted on a coverslip with 70\% glycerol and imaged with the miniscope. The miniscope's LED was used for reflectance imaging, while a broad-spectrum light source was used for transmission imaging. Imaging the same position before and after tartrazine will be required to draw stronger conclusions.

We applied transmission imaging to a full-thickness branch by removing the bark from the region of interest. Next, a hole was drilled through the branch from the bottom side, stopping 200 microns from the region of interest with a 1 mm drill bit. An optical fiber (Thorlabs FT400EMT) was then glued (Norland 65) into the pre-drilled hole. The illuminated site was imaged with light from the fiber and with the miniscope's LED. We present examples from two sections; one was treated with tartrazine and the other was not. Focus stacks were captured and projected with a complex wavelet transform projection.

\subsection{Parts List}
Table~\ref{tab:cwdt-parts-list} lists the main components used for the CWDT and tartrazine experiments.

\begin{table}[p!]
\centering
\small
\begingroup
\singlespacing
\renewcommand{\arraystretch}{0.95}
\caption{CWDT parts list.}
\label{tab:cwdt-parts-list}
\resizebox{\textwidth}{!}{%
\begin{tabular}{p{0.18\linewidth}p{0.18\linewidth}p{0.42\linewidth}p{0.12\linewidth}}
\toprule
Part & Make & Model & Price \\
\midrule
Microscope & Carson & MicroBrite & 18.99 \\
Optical Fiber & Thorlabs & FG050LGA, 50$\mu$m core 125$\mu$m cladding & 5.82 per meter \\
PLA Filament & Elegoo & 1.75 mm Black 1 KG & 13.99 \\
LED Flash Light & BLITZU & Gator 320 & 17.99 \\
Tartrazine dye & chemicalstore.com & FDC Yellow 5 Powder & 18.00 \\
Lapping sheets & Thorlabs & LF5P Silicon Carbide 5 $\mu$m Grit (10 Sheets) & 17.91 \\
Mirror Paint & Rust-Oleum & Mirror Effect Spray Paint 6 oz & 17.99 \\
Glycerol & 99.5\% glycerol/glycerin &  & 5.24 \\
Wood Plane & Lee Valley Tools & Veritas Miniature Bench Plane & 54.50 \\
Smartphone Camera & Motorola & ThinkPhone 50 MP f/1.8 & 270 \\
CPU & Intel & 11th Gen Core i7-11700F & 199.99 \\
GPU & NVIDIA & GeForce RTX 3060 & 250 \\
3D Printer & Bambu Lab & A1 Mini & 199.99 \\
\bottomrule
\end{tabular}%
}
\endgroup
\end{table}

\subsection{Results}
The following figures summarize the tartrazine tissue-clarification and transmission-imaging observations.

\begin{figure}[p!]
    \centering
    \includegraphics[width=4.5in]{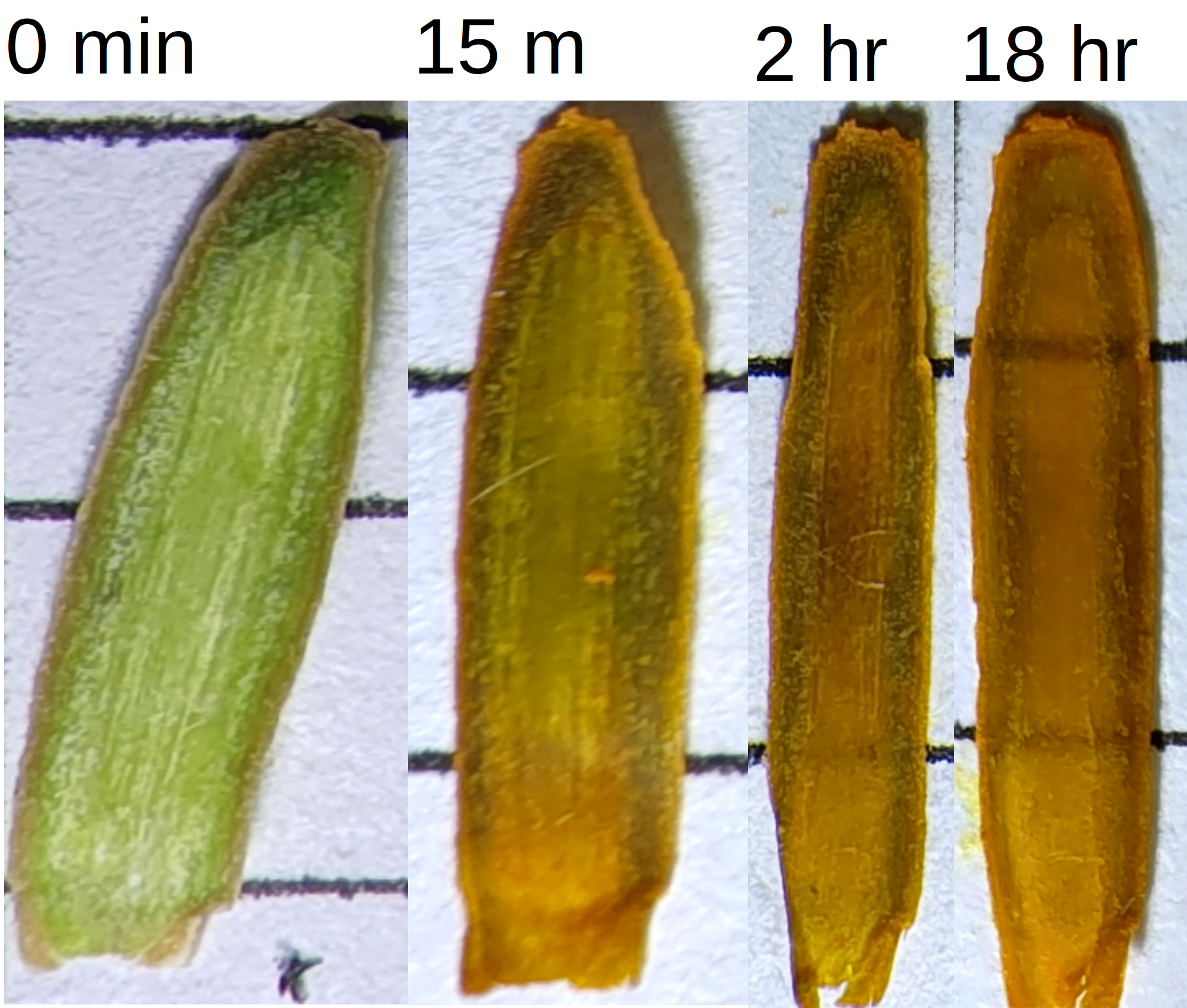}
    \caption{Poplar branch, sectioned longitudinally with 0.2 mm thickness and stained with tartrazine for up to 18 hours. Grid lines are spaced 5 mm apart.}
    \label{fig:cwdt-supp-plant}
\end{figure}

\begin{figure}[p!]
    \centering
    \includegraphics[width=4.5in]{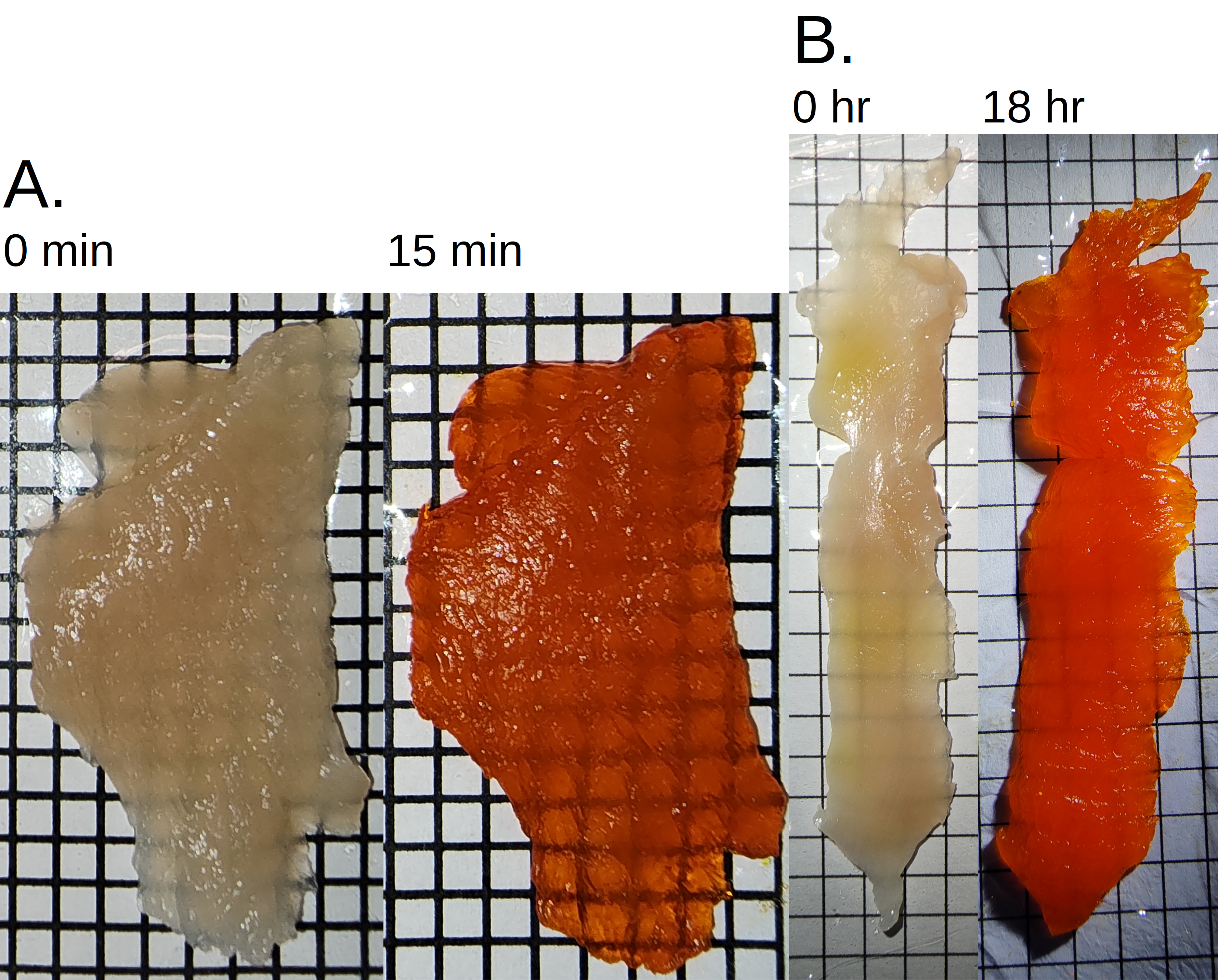}
    \caption{Raw chicken breast, sectioned with 0--3 mm thickness. A. Sample stained with tartrazine for 15 minutes; grid lines are 0.125 inches. B. Sample stained with tartrazine for 18 hours; grid lines are 5 mm.}
    \label{fig:cwdt-supp-chicken}
\end{figure}

\begin{figure}[p!]
    \centering
    \includegraphics[height=7.15in,keepaspectratio]{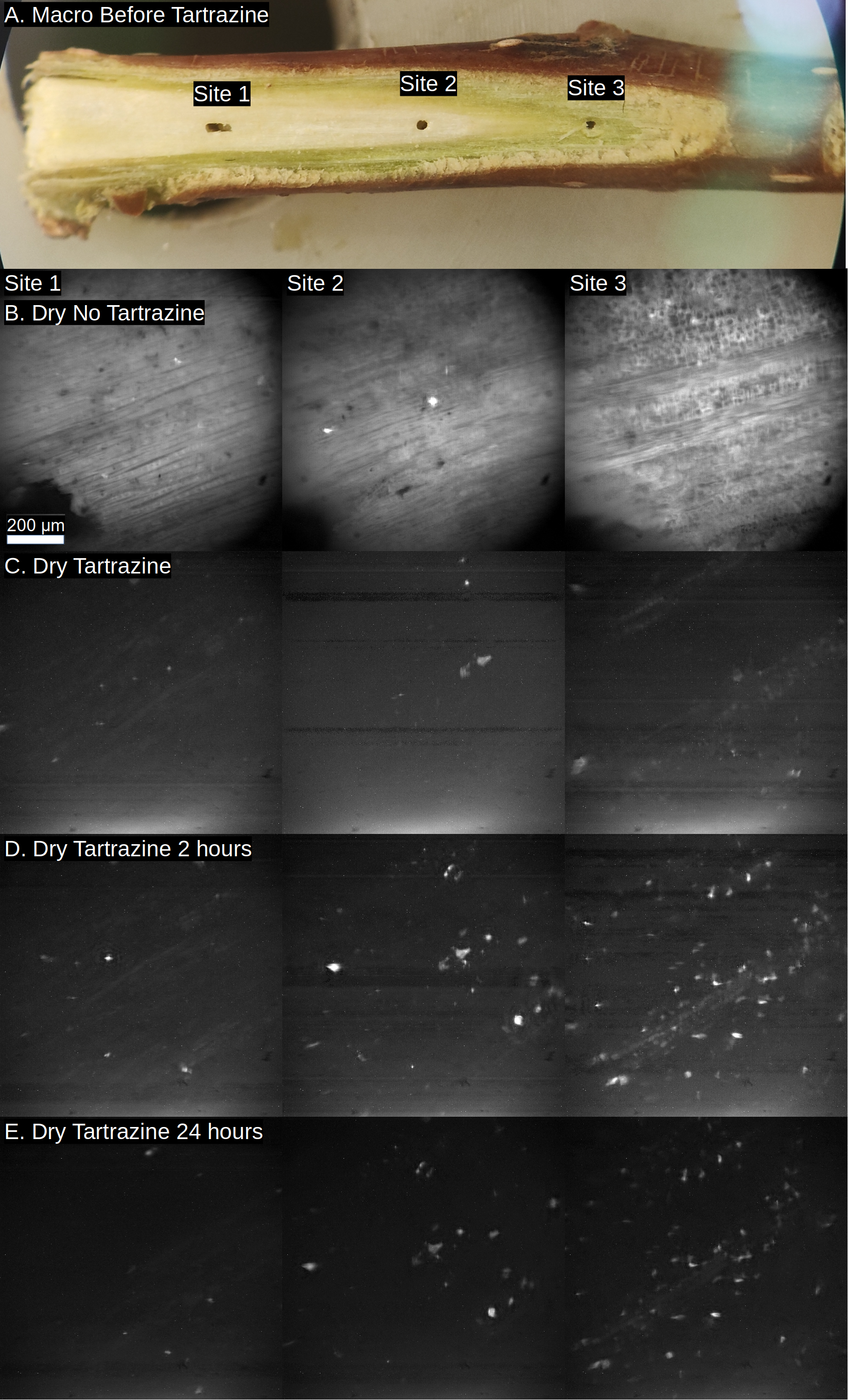}
    \caption{A poplar branch debarked, stained with tartrazine for 18 hours, and imaged with a UV autofluorescence miniscope. A. Macro image showing three fiducials indicating imaging sites. B. Control with no tartrazine and 50 ms exposure time. C. Tartrazine-stained branch, padded dry, with 500 ms exposure time. D. Tartrazine-stained branch after drying for 2 hours with 500 ms exposure. E. Tartrazine-stained branch after drying for 24 hours with 500 ms exposure. The scale bar is 200 microns.}
    \label{fig:cwdt-supp-miniscope1}
\end{figure}

\begin{figure}[p!]
    \centering
    \includegraphics[width=2.5in]{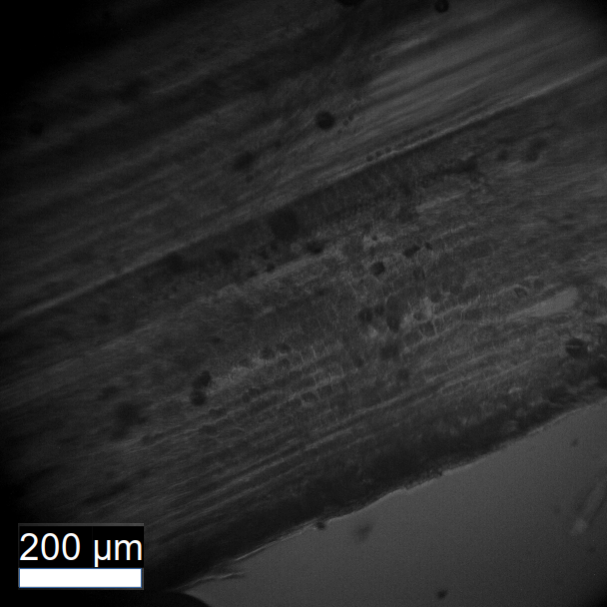}
    \caption{Transmission imaging of a 150 micron thick section of poplar wood soaked in tartrazine. The scale bar is 200 microns.}
    \label{fig:cwdt-supp-transmission}
\end{figure}

\begin{figure}[p!]
    \centering
    \includegraphics[width=3.5in]{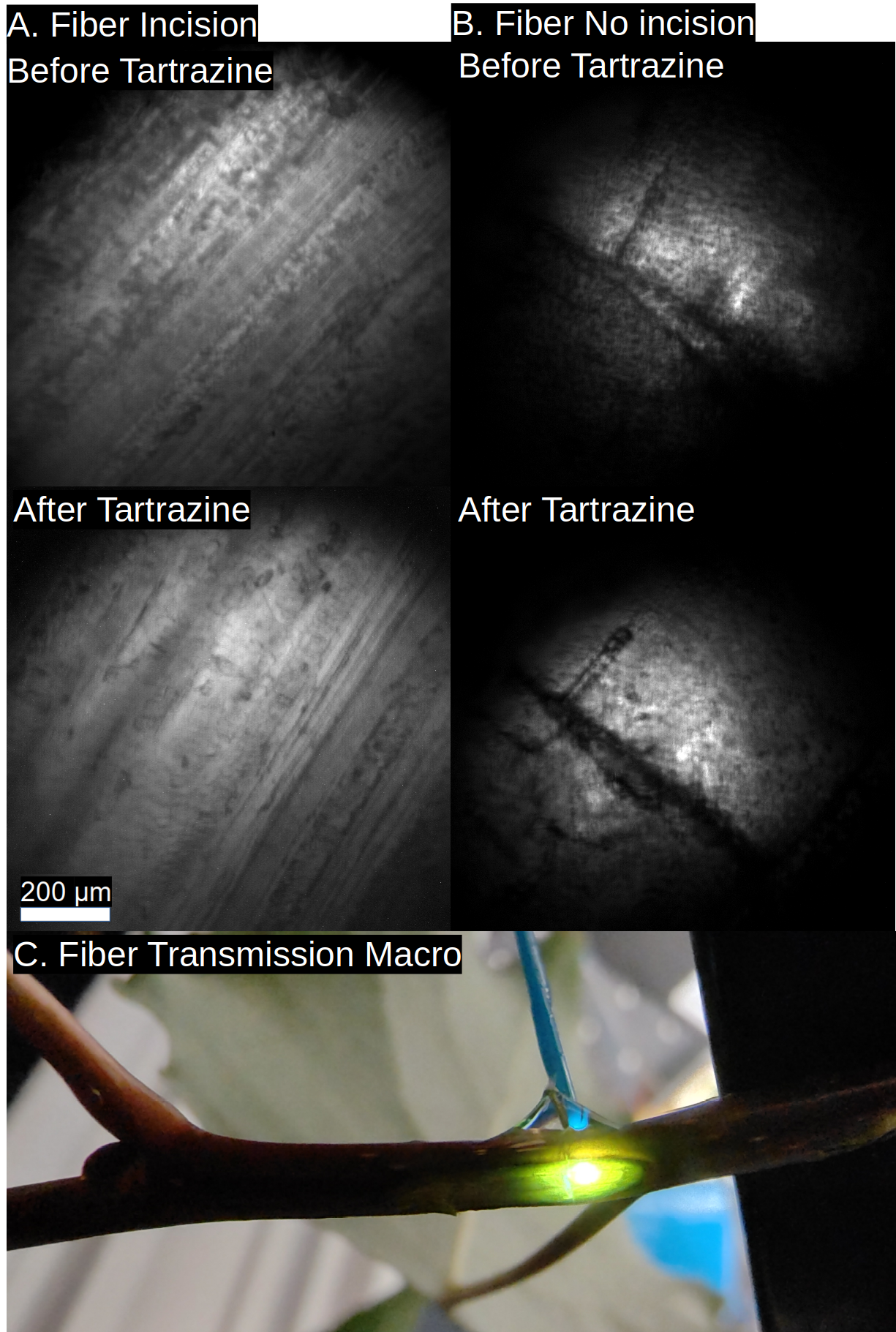}
    \caption{Transmission imaging of a poplar branch using an optical fiber inserted into a hole in the back of the branch before and after tartrazine application for 18 hours. A. Bark removed from the illuminated region. B. Bark not removed. C. Macro image of transmission illumination using an optical fiber. The scale bar is 200 microns for A and B.}
    \label{fig:cwdt-supp-fiber}
\end{figure}

\subsection{Gaussian Splatting}

An interesting alternative model for CWDT is a Gaussian splatting (GS) based method. Instead of modeling the tissue as a 3D grid of voxels, the sample can be modeled as a volume filled with translucent Gaussians, each with three position variables, three color variables, orientation, scale, shape, and opacity. Rather than designing a GS model from scratch, we modified open-source software \autocite{Kerbl2023,hbb1_torch_splatting_2025}.

The Gaussian splatting framework was adapted for continuous wave diffusive tomography (CWDT) simulation through key modifications to both the Gaussian model and rendering pipeline. The rendering system was transformed from uniform illumination to controllable point light sources suitable for CWDT simulation. Visible light source Gaussians were rendered at specified 3D positions with distance-dependent attenuation following the inverse square law. Automatic setting of the camera and light source positions was added. These modifications enabled simulation of multiple incident illumination positions with a fixed camera viewpoint. We tried two approaches, one with fixed Gaussian positions to better match the shape of the tissue with uniform Gaussian density, and one with learnable Gaussian positions to maximize MAE. For the fixed Gaussian model, the Gaussians were initialized in a grid and then random shifts were added to the positions of each Gaussian to eliminate the screen-door effect.

The models were trained for 50,000 iterations and achieved an MAE of 0.0939 for the original model and 0.1068 for the fixed-position model. For comparison, the voxel model achieved an MAE of 0.052. While the GS model produced a plausible tomographic reconstruction of the tissue, it performed significantly worse in capturing fine sample detail than the main voxel-based model. We were limited to around 12,000 Gaussians by the VRAM limits of an NVIDIA GeForce RTX 3060 with 12 GB. It may be possible to exceed the quality of the voxel-based model with sufficient optimization and a sufficiently powerful computer, but the voxel model produced a significantly higher-quality result with only 2 GB of VRAM.

\begin{figure}[p!]
    \centering
    \includegraphics[width=\linewidth]{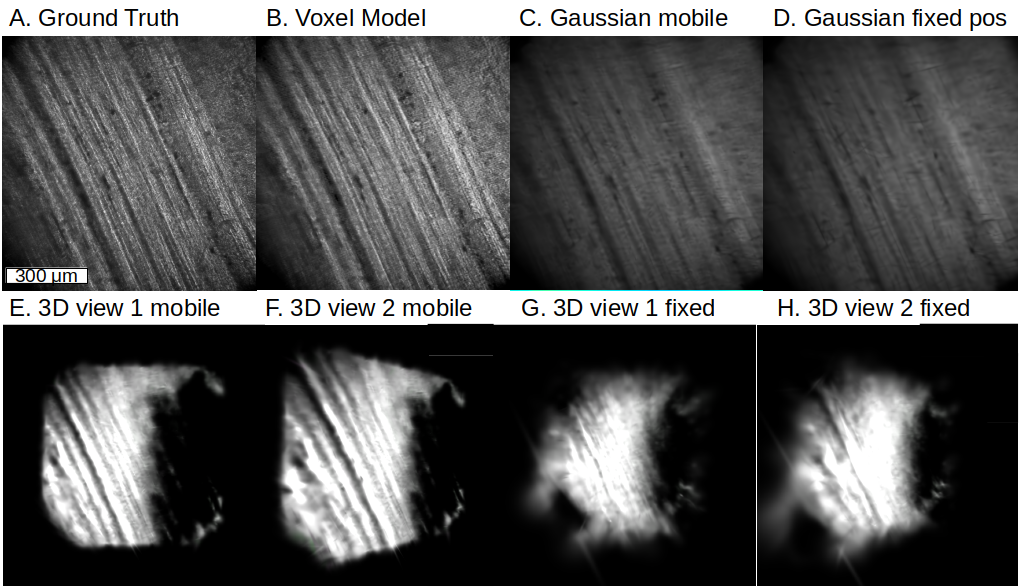}
    \caption{Comparison of voxel-based and Gaussian-splatting CWDT reconstructions. A. Ground-truth image. B. Example from the voxel-based light transport model. C. Gaussian splatting model with variable Gaussian positions. D. Gaussian splatting model with fixed Gaussian positions. E-F. Two rotated views of the variable-position Gaussian model. G-H. Two rotated views of the fixed-position Gaussian model. The example image is grayscale because it was captured with the V4 UCLA miniscope, an early version of the CWDT setup.}
    \label{fig:cwdt-supp-gaussian}
\end{figure}

The voxel and Gaussian-splatting outputs are compared in Figure~\ref{fig:cwdt-supp-gaussian}.

\subsection{Conclusion}

We successfully replicated the results of the original tartrazine paper with chicken breast imaged on top of graph paper (Figure~\ref{fig:cwdt-supp-chicken}). We found improved tissue clarity in poplar wood measured on top of graph paper (Figure~\ref{fig:cwdt-supp-plant}). It should be noted that dryness and surface reflectance of the samples were not adequately controlled using a coverslip and glycerol mounting medium, reducing transmission. Nevertheless, we proceeded with UV autofluorescence and transmission experiments.

We explored UV autofluorescence imaging of lignin in wood before and after treating samples with tartrazine. We found that fluorescence levels change as the sample dries, so we exposed one sample to air (Figure~\ref{fig:cwdt-supp-miniscope1}) and tested a separate immersion-oil-covered sample. Tartrazine significantly decreased fluorescence intensity, necessitating an increase in exposure time from 50 ms to 500 ms. We found no evidence that tartrazine application led to imaging depths below the material's surface with UV autofluorescence imaging of lignin at the wavelengths used here. These frequencies are strongly absorbed by tartrazine. In future research, staining with Nile Red may have better results because its excitation and emission bands better avoid tartrazine's 300--500 nm absorption.

Transmission imaging of poplar samples treated with tartrazine showed some promising results. Thin sections treated with tartrazine produced detailed images revealing the structure of the wood (Figure~\ref{fig:cwdt-supp-transmission}). Unfortunately, thin sections are incompatible with imaging living cells. We therefore experimented with transmission illumination of living branches using an optical fiber inserted into a hole drilled in the back of the sample (Figure~\ref{fig:cwdt-supp-fiber}). When the bark was removed from the sample, tartrazine treatment resulted in vessel elements that were slightly more defined. When the bark was not removed, tartrazine did not make a significant difference, and we were only able to visualize the bark surface. In future experiments, movement of the optical fiber illumination could be used to calculate a 3D map of the branch using continuous-wave diffuse optical tomography \autocite{Culver2003}.

\bibliographystyle{unsrtnat}
\bibliography{references}

\end{document}